\documentclass[graybox]{svmult}

\usepackage{type1cm}        
\usepackage{makeidx}         
\usepackage{graphicx}        
\usepackage{multicol}        
\usepackage[bottom]{footmisc}

\usepackage[fencedCode]{markdown}

\usepackage[newfloat=true]{minted}

\usepackage{caption}

\usepackage{xspace}

\usepackage{wrapfig}

\usepackage{hyperref}

\usepackage{cleveref}

\usepackage{newtxtext}       %
\usepackage[varvw]{newtxmath}       

\newcommand{\pelelm}{{PeleLM}\xspace}

\newcommand{\exasgd}{{ExaSGD}\xspace}
\newcommand{\opencarp}{{openCARP}\xspace}
\newcommand{\openfoam}{{OpenFOAM}\xspace}
\newcommand{\ogl}{{OGL}\xspace}

\newcommand{\petsc}{{PETSc}\xspace}
\newcommand{\trilinos}{{Trilinos}\xspace}
\newcommand{\hiop}{{HiOP}\xspace}
\newcommand{\sundials}{{Sundials}\xspace}
\newcommand{\ginkgo}{{Ginkgo}\xspace}

\makeindex             

\begin{document}

\setminted{
    fontsize=\small  
}
\newmintinline{cpp}{}

\title*{Software Development Aspects of Integrating Linear Algebra Libraries}
\author{Marcel Koch\orcidID{0009-0004-8333-9313} and\\ Tobias Ribizel\orcidID{0000-0003-3023-1849} and\\ Pratik Nayak\orcidID{0000-0002-7961-1159} and\\ Fritz G\"{o}bel and\\ Gregor Olenik\orcidID{0000-0002-0128-3933} and\\ Terry Cojean\orcidID{0000-0002-1560-921X}}
\institute{Marcel Koch \at KIT, Scientific Computing Center (SCC), Zirkel 2, 76131 Karlsruhe, Germany, \\ \email{marcel.koch@kit.edu}
\and Tobias Ribizel \at TUM, Computational Mathematics, Bildungscampus 3, 74076 Heilbronn, Germany \\ \email{tobias.ribizel@tum.de}
\and Pratik Nayak \at KIT, SCC, \email{pratik.nayak@kit.edu}
\and Fritz G\"{o}bel \at KIT, SCC, \email{fritz.goebel@kit.edu}
\and Gregor Olenik \at KIT, SCC, \email{gregor.olenik@kit.edu}
\and Terry Cojean \at KIT, SCC, \email{terry.cojean@kit.edu}}
\authorrunning{Marcel Koch et al}
%
%
\maketitle

\abstract*{
Many scientific discoveries are made through, or aided by, the use of simulation software.
These sophisticated software applications are not built from the ground up, instead they rely on smaller parts for specific use cases, usually from domains unfamiliar to the application scientists.
The software library Ginkgo is one of these building blocks to handle sparse numerical linear algebra on different platforms.
By using Ginkgo, applications are able to ease the transition to modern systems, and speed up their simulations through faster numerical linear algebra routines.
This paper discusses the challenges and benefits for application software in adopting Ginkgo.
It will present examples from different domains, such as CFD, power grid simulation, as well as electro-cardiophysiology.
For these cases, the impact of the integrations on the application code is discussed from a software engineering standpoint, and in particular, the approaches taken by Ginkgo and the applications to enable sustainable software development are highlighted.
}
\abstract{
Many scientific discoveries are made through, or aided by, the use of simulation software.
These sophisticated software applications are not built from the ground up, instead they rely on smaller parts for specific use cases, usually from domains unfamiliar to the application scientists.
The software library Ginkgo is one of these building blocks to handle sparse numerical linear algebra on different platforms.
By using Ginkgo, applications are able to ease the transition to modern systems, and speed up their simulations through faster numerical linear algebra routines.
This paper discusses the challenges and benefits for application software in adopting Ginkgo.
It will present examples from different domains, such as CFD, power grid simulation, as well as electro-cardiophysiology.
For these cases, the impact of the integrations on the application code is discussed from a software engineering standpoint, and in particular, the approaches taken by Ginkgo and the applications to enable sustainable software development are highlighted.
}

\section{Introduction}
\label{sec:intro}

Integrating numerical linear algebra libraries has considerable impact for application software.
They provide new functionality, or optimize existing ones.
Typically, application developers have little expertise in numerical linear algebra.
Thus, through the integrations, the application software can benefit from functionality developed by experts in this field.
The integrations come at a cost, though.
They require development time, which is usually scarce for application scientists.

Both application and library developers have to ensure that the development cost stays minimal.
This paper considers efforts within the numerical linear algebra library \ginkgo, and selected applications, with integrations of it, taken to minimize development and maintenance effort.
This paper will first give a brief overview of the related work in \cref{sec:related}.
Then, in \cref{sec:ginkgo} \ginkgo's main concepts regarding integrations are presented.
A corresponding view on integration from the application side is discussed next in \cref{sec:integration}.
Several case studies are presented in \cref{sec:cases}.
These highlight the key points from the two previous sections.
Finally, a summary of the main points is provided in \cref{sec:summary}.
\section{Related work}
\label{sec:related}

Many software libraries provide numerical linear algebra functionality.
Some of the most versatile libraries are \petsc~\cite{petsc-user-ref} and \trilinos~\cite{trilinos}.
Both offer features for linear algebra, but they also encompass more areas, such as non-linear solver, mesh handling, parallel communication, or discretization frameworks.
As such, they are attractive to a variety of applications.
The \petsc user manual, for example, has more than 4000 citations on Google Scholar and \trilinos close to 1500\footnote{From \url{https://scholar.google.com/} accessed on 2024.05.06}.
Most of these are application papers that usually present only the results of integration of one of these libraries, and don't discuss their approach to the integration in detail.
Papers on the library themselves, on the other hand, e.g. \petsc's AMG implementation~\cite{petsc-amg} or an overlapping Schwarz method in \trilinos~\cite{trilions-frosch}, often focus on new algorithmic improvements, leaving out again details on the integration aspects.

In~\cite{dongarra-evolution} and~\cite{hsl-history} the history of important numerical software libraries is presented.
Dongarra's Turing lecture~\cite{dongarra-evolution} focuses on the developments of LAPACK and adjacent packages, e.~g. BLAS, ScalLAPACK, or SLATE, while~\cite{hsl-history} discusses the mathematical software library HSL.
Both articles show how libraries can adapt to emerging needs of the community to stay relevant.
Since the articles present the rich history of longstanding libraries, little place is left for technical details on integration.
Nevertheless, Dongarra also notes the funding and personnel cost for maintaining such long-standing efforts.

The performance impacts of the integrations in \cref{sec:cases} have already been published.
\ginkgo's batched solver, which provides most of the performance benefits for \pelelm, are described in~\cite{aggarwalBatchedSparseIterative2021,aggarwalPreconditionersBatchedIterative2022}.
The paper~\cite{exasgd} shows the performance of \ginkgo's direct solver in the context of the \exasgd application.
Additionally, the summary paper~\cite{ijhpc-ginkgo} provides an overview of the performance gains of multiple applications from the Exascale Computing Project (ECP)\footnote{\url{https://www.exascaleproject.org/}} that use \ginkgo, including \pelelm and \exasgd.
The \openfoam results using \ogl are available in \cite{olenik2024Towards}.
All of these papers, again, provide little consideration to the integration effort.
Only the papers on \ogl are an exception since they discuss the design of \ogl which impacts the integration.

\section{The \ginkgo Library}
\label{sec:ginkgo}


\ginkgo is a software library for sparse numerical linear algebra.
It started as part of the ECP with the goal of providing an efficient and portable library for heterogeneous hardware.
In particular, \ginkgo runs on all major GPU architectures.
The respective backends are written in their native language to gain the most performance, i.e. CUDA for NVIDIA GPUs, HIP for AMD GPUs, and SYCL for Intel GPUs.
On the algorithmic side, the main focus lies on iterative and direct solvers for sparse linear systems.
Together with preconditioners and reorderings, \ginkgo is capable of solving complex problems efficiently on a variety of different architectures.
This section introduces the core concepts of \ginkgo and discusses how they facilitate the integration of \ginkgo into application software.

\subsection{Concepts}
\label{sec:ginkgo:concepts}

At its core, \ginkgo is a modern C++ library that values performance, flexibility, and sustainability.
The aforementioned hardware specific backends offer excellent performance through optimized low-level kernels, while the clear type hierarchy offers flexibility in terms of functionality for the users.
Through modern C++ patterns (requiring the standard C++14), the library provides a sustainable design.
Finally, exhaustive testing and benchmarking ensure the correctness of the library.

\paragraph{Kernel Backends}
The library is split into a core library and the different backend libraries.
Performance critical kernels are implemented in the backends, and the higher level algorithmic structure is in the core library.
The separation enables Ginkgo to directly target the hardware in question.
This allows Ginkgo to easily implement complex algorithms on an abstract level, while gaining performance through hardware-optimized kernels.
Compatibility layers, such as Kokkos~\cite{carter_edwards_kokkos:_2014}, typically add some overhead compared to native implementations as shown in~\cite{p3hpc-1}, which \ginkgo circumvents.

The separation also benefits encapsulation within \ginkgo.
Any dependencies on the vendor specific languages are restricted to their respective backend and not propagated through the whole library.
An application thus only needs a standard C++ compiler to compile with \ginkgo, even if the application will use for example \ginkgo's CUDA kernels.

The backend design not only offers performant kernels, it also provides flexibility in terms of the used hardware.
\ginkgo can be built with all backends enabled at the same time.
This makes it not a compile-time choice, which hardware should be used and which not, but a runtime choice instead.
Which backend is chosen only depends on the implementation of the \cppinline{Executor} interface used at runtime.
The following example shows how different executors are used, depending on a runtime parameter:
\begin{minted}{c++}
void app_run(std::shared_ptr<gko::Executor> exec, ...);
// ...
{
  auto exec = cuda ? gko::CudaExecutor::create(...)
                   : gko::OmpExecutor::create(...);
  app_run(exec, ...);
}
\end{minted}
Any \ginkgo operation on \cppinline{app_run} will then use the kernels belonging to the dynamic type of \cppinline{exec}.
Furthermore, as \cppinline{app_run} only depends on the abstract \cppinline{Executor} interface, it is not necessary to recompile this function when switching between executors.

\paragraph{Linear Operators}

The \ginkgo library is mainly designed around the linear operator concept.
The type \cppinline{LinOp} represents an abstract interface for the mathematical notion of a linear operator.
This encompasses of course matrices, but also preconditioners and linear solvers, although they satisfy linearity only up to the achieved solving precision.
All \ginkgo types implementing these concepts are implementations of the abstract \cppinline{LinOp} type.
The primary requirement is that the implementations provide a mapping between vector spaces.

The abstract linear operator concept offers great flexibility for choosing different algorithms.
Through polymorphism, different implementations of, for example, matrix formats can be used interchangeably in algorithms that only rely on the operator application.
This is true for most Krylov methods.
So, users can easily switch out the matrix format to be used in a Krylov solver, and pick the format that is most suitable for their specific application.
However, the flexibility is not restricted to matrix formats.
The Krylov (and direct) solvers themselves, as well as the preconditioners are also implementations of the \cppinline{LinOp} interface.
Thus, users can easily try out a wide range of functionality, for example switching between applying a matrix or its inverse approximated by a solver:
\begin{minted}{c++}
void operate(const gko::LinOp*, ...);
// ...
{
  if(invert){
    auto solver = gko::solver::Cg<>::build()->generate(mtx);
    operate(solver.get(), in, out);
  } else {
    operate(mtx.get(), in, out);
  }
}
\end{minted}

\paragraph{Design Patterns}

\ginkgo uses design patterns to simplify the code, convey intent, and provide safe code.
A key pattern is the use of smart pointers.
These enable behavior close to automatic garbage collection, while still having completely deterministic behavior.
Through them, the lifetime and ownership of polymorphic objects is managed.
Either an object only has a single owner, then a \cppinline{std::unique_ptr} keeps the object alive as long as the pointer stays in scope.
The following gives a short example using one of \ginkgo's matrix types:
\begin{minted}{c++}
{
  std::unique_ptr<gko::LinOp> mtx = gko::matrix::Csr::create(exec, ...);
  // operate on mtx
  // ...
} // after leaving the scope mtx is automatically deleted
\end{minted}
Alternatively, an object might have multiple owners, e.g. a matrix that is owned by a solver and a preconditioner.
In this case, a \cppinline{std::shared_ptr} keeps the object alive while at least one owner stays in scope.
Once all owners are out-of-scope, the object is finally deleted.
This creates a semi-automatic garbage collector behavior that prevents common memory issues, such as leaks, or double frees, while using a simple API with deterministic behavior and minimal overhead.

Lastly, \ginkgo is designed to offer great compatibility with other libraries through its transparent API.
Part of the transparent API is to provide access to the stored data of the previously mentioned data structures through raw pointers.
In contrast to smart pointers, these don't modify the ownership, but this introspection of the data layout allows other libraries to operate directly on data owned by \ginkgo.
Compatibility in the other direction is also supported, since most data structures can be created by passing in raw pointers to existing data, resulting in \ginkgo objects that don't take ownership of the data.
Both options are presented in the following snippet:
\begin{minted}{c++}
// extract data (and number of non-zeros) from a Ginkgo CSR matrix
double* values = mtx->get_values();
gko::size_type nnz = mtx->get_num_stored_elements();

// lend data to Ginkgo
double* app_ptr = ...; // given from application
auto view = gko::make_array_view(exec, app_size, app_ptr);
\end{minted}

\subsection{Facilitating Integrations}
\label{sec:ginkgo:integration}

The aforementioned design concepts provide distinct benefits when integrating \ginkgo.
Besides offering high-performance kernels, the runtime selectable backends provide great flexibility when interacting with other realizations of platform portability.
Many portability approaches are configurable only at compile-time.
This includes portability layers like Kokkos.
An application that uses such a compile-time approach has to make sure that all components use the same configured backend.
Since the backend choice for \ginkgo is delayed until runtime, it's easy to ensure that it uses the correct backend.

Furthermore, \ginkgo's type hierarchy, especially the \cppinline{LinOp} interface, offers
a flexible and extensible API.
Not only can existing Ginkgo algorithms be switched out easily, applications can also provide their own implementation of \ginkgo's interface and then use their types in \ginkgo's algorithms.
For example, this allows users to implement an operator for matrix-free stencil applications.
The matrix-free operator can then be used in \ginkgo's Krylov solvers.
Thus, applications gain flexible access to algorithms through \ginkgo, with a simple extension interface to inject specialized operations.

Finally, \ginkgo's transparent API greatly helps the data exchange with application code.
By exposing raw pointers, an application can easily pass \ginkgo data, such as vectors, to other components without them relying also on \ginkgo, while the ownership stays with \ginkgo.
Additionally, data that is owned by the application can be lent to be used within \ginkgo.
This helps the application to create clear boundaries between their internal components without requiring expensive copies.
\section{Integration Approaches in Application Software}
\label{sec:integration}

The previous section presented approaches within \ginkgo to ease the effort of integrating it into an application.
Now, the application side of the integration is considered.
In general, an application can follow two different approaches to the integration.
First, the application can use tight coupling with \ginkgo.
The other option is loose coupling.
These approaches are not specific to \ginkgo, but rather common approaches in the design of software applications.
They reflect how deeply intertwined the application becomes with a library.

The strongest intertwining is the tight coupling.
In this case, the application uses the \ginkgo data structures directly and might even use them outside the linear solver setting.
Loose coupling leads to less intertwining by relying on abstract interfaces instead of \ginkgo directly, which can be considered as an application of the facade design pattern as mentioned in \cite{big4}.
\ginkgo supports both modes, and the examples in \cref{sec:cases} show different variations on them.
The \openfoam example uses tight coupling, and \opencarp transitioned from tight coupling with \petsc to loose coupling for both \petsc and \ginkgo.
Both \exasgd and \pelelm use loose coupling, although in different variations.

\subsection{Tight Coupling}
\label{sec:integration:tight}

\ginkgo provides the basic data structures necessary for linear algebra.
Both dense vectors and sparse matrices with different formats are available to use as the backbone of an application's linear algebra component.
Additionally, \ginkgo offers a storage type for one-dimensional arrays.
The array type is directly linked to the \cppinline{Executor} concept, which allows allocating memory on different devices besides the CPU.
Together with simple APIs for copying data between arrays of different executors, \ginkgo provides the means for arbitrary storage using different devices.

In contrast to other popular numerical linear algebra packages, \ginkgo's scope is limited to linear algebra.
Applications might require other concepts that are not strictly part of linear algebra, mesh handling for example.
Longstanding libraries such as \petsc or \trilinos provide these additional functionalities.
Although \ginkgo is more focused than those, it might still be used as the sole linear algebra backend.
Since \ginkgo allows for direct data exchange without overhead, it can easily be used in conjunction with other libraries that cover the missing areas.

The Example in \Cref{lst:tight_coupling} shows how an application could implement an implicit Euler time stepper by using a tight coupling with \ginkgo.
The first thing to note is that the application class stores \ginkgo objects directly.
The \cppinline{advance} function then uses these objects and calls methods directly on them.
Additionally, as the function expects \ginkgo objects as parameters, the calling code must also use \ginkgo objects directly.
Thus, the dependency of \ginkgo extends further than this class.

\begin{listing}[htbp]
\begin{minted}{c++}
using Dense = gko::matrix::Dense<>;

class ImplicitEuler{
public:
  void advance(Dense* u_n, Dense* dx){
    // assemble right-hand-side f_ based on u_n

    Ainv_->apply(f_, x_);

    u_n->add_scaled(dx, x_);
  }

private:
  std::unique_ptr<Dense> x_;
  std::unique_ptr<Dense> f_;
  std::unique_ptr<gko::LinOp> Ainv_;
}
\end{minted}
\captionof{listing}{A tight-coupling application code example that implements an Euler time-stepper using Ginkgo functionality.}
\label{lst:tight_coupling}
\end{listing}

The example highlights some of the advantages and disadvantages of the tight coupling.
As \ginkgo types are used directly, no additional type hierarchy on the application side is necessary.
But the implementation on the application side now deeply depends on \ginkgo.
Generally speaking, implementations using tight coupling have their lifetime tied to the integrated libraries.
If the integrated library stops receiving updates, the application is stuck with potentially out-of-date code.
Either the application would have to update the library or replace it.
Both would be major undertakings for the application.

Another not yet mentioned benefit is the simpler dependency management.
The application always depends on the integrated library, so no special handling has to be developed if a library is missing.
This will usually simplify the application's build system.

\subsection{Loose Coupling}
\label{sec:integration:loose}

Compared to tight coupling, loosely coupled integrations will use \ginkgo only through indirection.
First, the application needs abstract interfaces for its linear algebra component.
The interfaces represent the boundary between the application code and the external libraries, such as \ginkgo.
Interface code is called with application data, and expects application data as output if applicable.
But internally, implementations of the interface will use \ginkgo types and functions.
This might incur some overhead due to copying.
Although both sides of the interface use the same conceptual representation, e.g. a matrix in CSR format, the underlying data layouts might differ, which requires copies.
On the other hand, if the underlying data layouts are the same, no copy should be necessary.
Thus, if libraries can communicate the used data layouts through their APIs, then interoperability can be improved through the elimination of copies, as will be seen in the case studies from \cref{sec:cases:ecp}.

In the context of linear algebra, these interfaces usually involve vectors, matrices, and solvers.
These interfaces can provide multiple purposes.
The solver interface provides the solution to a linear system, and the vector and matrix interface can provide conversions between the application's native format and \ginkgo, or other features such as assembly.
Applications might provide multiple implementations of an interface for \ginkgo.
If they differentiate between Krylov solvers on a type level, for example, then multiple implementations of the solver interface might be necessary to cover different \ginkgo solvers.

The interfaces might also provide simpler access to complex functionality.
A solver interface could offer a method to regenerate the solver with a new matrix.
The \ginkgo implementation for this would contain multiple lines of code, with different \ginkgo functionality used, because \ginkgo doesn't provide the regenerate functionality through a simple API.
Thus, the interface can hide complexity in \ginkgo, making it easier to use for the application specific purposes.

It is not necessary to provide all the mentioned interfaces.
If only the solution to a linear system is required, without additional matrix or vector algebra, then the solver interface would be enough.
The conversions between application and \ginkgo data are then handled purely internally.
Ultimately, it's up to the application if all three interfaces are necessary, or if a subset suffices.

The example in \Cref{lst:loose_coupling} shows an implementation of a contrived abstract solver interface.
In this case, the application only provides the solver interface, so the in- and outputs are defined as application types.
During the constructor, an application matrix is provided.
This is converted into a \ginkgo matrix using an explicit copy, since it is assumed that different data layouts are used.
For the solution method, application vectors are provided, but no copy is performed.
Now, the assumption is that both \ginkgo and application vectors have the same data layout.
This is typically easier to guarantee since \ginkgo uses a contiguous data layout with strides, which is widely used, e.g. in the BLAS API, or for Kokkos Views.

The only \ginkgo specific type in the interface is the executor \cppinline{exec}.
If an application is already using different execution spaces, e.g. CPU execution, and GPU execution, then a mapping between the applications' notion of execution spaces and \ginkgo's executors has to be defined.
In the special case of Kokkos' execution spaces, \ginkgo already provides these mapping.
For applications that were not using GPUs before, this might consist in mapping \cppinline{std::strings} to \ginkgo's executors.

\begin{listing}[p]
\begin{minted}{c++}
class GinkgoCG: public AbstractSolver{
  using Dense = gko::matrix::Dense<>;
  using Csr = gko::matrix::Csr<>;
  using Cg = gko::solver::Cg<>;
  
public:
  void solve(const AppVector& b, AppVector& x) override
  {
    // Create views on the application data.
    // If the application data uses host memory, then instances of exec_
    // needs to be replaced by exec_->get_master().
    auto gko_x = Dense::create(
                    exec_, gko::dim<2>{b.num_rows(), b.num_cols()},
                    gko::make_array_view(exec_, b.total_size(), b.data()),
                    b.stride());
    // This is the const-correct version of the call above.
    auto gko_b = Dense::create_const(
                    exec_, gko::dim<2>{b.num_rows(), b.num_cols()},
                    gko::make_const_array_view(exec_, b.total_size(),
                                               b.data()),
                    b.stride());
    solver_->apply(gko_b, gko_x);
    // since views are used, the AppVector x already contains the result
    // otherwise an extra copy is necessary
  }
  
  GinkgoCG(std::shared_ptr<const gko::Executor> exec, AppMatrix& A,
           int max_it, double eps)
    : AbstractSolver(), exec_(exec)
  {
    gko::matrix_data<> md(gko::dim<2>{A.num_rows(), A.num_cols()};
    //Fill md by extracting the triplet (row-index, col-index, value) for 
    // each non-zero in A and pass it to 
    // md.nonzeros.emplace_back(row, col, value)
    auto gko_A = Csr::create(exec);
    gko_A->read(md);
    solver_ = Cg::build()
                .with_criteria(gko::stop::Iteration::build()
                                 .with_max_iters(max_it),
                               gko::stop::ResidualNorm<>::build()
                                 .with_reduction_factor(eps))
                ->generate(std::move(gko_A));
  }
  
private:
  std::shared_ptr<const gko::Executor> exec_;
  std::unique_ptr<Cg> solver_;
};    
\end{minted}
\captionof{listing}{A loose-coupling application integration example, implementing a GinkgoCG solver by following an abstract solver interface.}
\label{lst:loose_coupling}
\end{listing}

Compared to the tight coupling, loose coupling exhibits advantages that support the applications' freedom of choice.
As the dependency is hidden through an interface, it's easy to change, remove, or add external libraries implementations.
Furthermore, the dependency is restricted to a clear set of implementation files.
Another benefit of the interfaces is the restriction of the integrated library to only the necessary functionality.
On the other hand, the interface introduces an additional class hierarchy to manage, as well as boilerplate code in some instances.
Also, managing the dependencies in the build system is more complex.
\section{Case Studies}
\label{sec:cases}

In the following section, examples are brought up to highlight the improvements applications can gain from integrating \ginkgo, as well as the differences in their integration approaches.
First, two ECP applications, \exasgd and \pelelm, are discussed.
Both use loose coupling to integrate \ginkgo.
While \exasgd benefits from the direct solvers in \ginkgo, \pelelm exploits the batched iterative solvers.
The third application is \opencarp, which transitioned from a tight coupling with \petsc to a loose coupling with both \petsc and \ginkgo.
It benefits from the enabled GPU acceleration.
Finally, the \ogl library uses tight coupling to provide \ginkgo solvers for \openfoam.
Again, through these solvers, GPU acceleration is enabled for parts of computational fluid dynamics (CFD) simulations using \openfoam.

\subsection{ECP Applications}
\label{sec:cases:ecp}

Both the \exasgd and \pelelm applications benefited greatly from their integration with \ginkgo.
As the major findings have been presented in the papers~\cite{aggarwalBatchedSparseIterative2021,aggarwalPreconditionersBatchedIterative2022,exasgd}, and additionally in the summary paper~\cite{ijhpc-ginkgo}, the performance gain for the applications are not discussed here.
Instead, the focus lies on the integration approaches.
The summary paper also mentions integrations with MFEM, and XGC.
These are not followed up on here, since XGC is very similar to \pelelm in how \ginkgo is used, and MFEM doesn't classify as a science application, since it is again a library that other applications will integrate.

The \textbf{\exasgd} project simulates the power grid of the US in order to make projections for future developments.
At its core, it solves non-linear optimization problems, which is done through \hiop~\cite{hiop_techrep}.
The resulting linear systems tend to be highly ill-conditioned, requiring the use of a direct solver.
In addition, these systems have little structure, which makes prominent direct solvers such as SuperLU less effective.
A new direct solver has been co-developed within \ginkgo to address these issues.
The authors provide a detailed overview in~\cite{ginkgo-direct}.
By integrating \ginkgo into \hiop, platform portable direct solvers became available, as discussed in~\cite{exasgd}.

As mentioned earlier, \exasgd (through \hiop) integrates with \ginkgo through loose coupling.
More specifically, it only provides a solver interface for \ginkgo to implement.
Thus, the integration requires a few lines of code, in total about 470 lines.
Together with a few outside changes, e.g. to the build system or code examples, the initial implementation was done in 700 line of code (LOC) changes\footnote{\url{https://github.com/LLNL/hiop/pull/433/}}.
This marks a relatively low burden on \hiop's code base.

The integration is very similar to the example provided in \cref{sec:integration:loose}.
In both cases, an application matrix is copied over to a \ginkgo matrix during initialization, while views on application vectors are used in the solver application phase.
Additionally, more application-specific interfaces are provided.
These concern updating the values of a matrix, without changing the structure, and recomputing the factorization.
Since the direct solver classes in \ginkgo don't provide this functionality directly, the implementation complexity is hidden by the interface.

During the interface initialization, the \ginkgo objects can be customized.
Since \hiop only requires a specific subset of \ginkgo, the customization is rather limited.
Only five options are provided in a mapping from \hiop options to \ginkgo customization.
Besides that, many \ginkgo settings are fixed.
The solver is always an LU solver, possibly wrapped in GMRES depending on the options, and MC64 is always used as an apriori reordering of the system.

The \textbf{\pelelm} application~\cite{pelelm} is used for CFD simulation, specifically for reactive flow in the Low Mach number regime.
The discretization approach requires solving many independent ODEs.
These are solved using \sundials~\cite{sundials}, which requires the solution of many small independent linear systems.
To solve these efficiently, batched iterative solvers have been developed in \ginkgo.
The papers~\cite{aggarwalBatchedSparseIterative2021,aggarwalPreconditionersBatchedIterative2022} offer more details on the performance aspects.

Although the \pelelm integration through \sundials shares many similarities to the \hiop integration, it had a vastly different starting point.
In contrast to \hiop, \sundials is mainly a C code.
This introduces additional friction, as many modern C++ design principles which \ginkgo uses are not compatible with C design patterns.
Integrating Gingko using a C API would lead to struggles due to C++ features that are not part of C, mostly RAII-based object management and templating.
To alleviate the differences, \sundials introduced a new C++ API instead.
This API takes advantage of templating to let users decide which concrete \ginkgo type should be used in the implementation of the interface.

The extra work is reflected in the integration effort.
About 8000 LOC additions and 2500 LOC deletions were necessary in the initial pull request\footnote{\url{https://github.com/LLNL/sundials/pull/205}} (PR), which included the new C++ API.
Despite the large-scale API changes, the changes solely related to \ginkgo are in the same order as \hiop.
The \ginkgo implementation of the linear solver C++ API requires about 350 LOC.
In addition, the matrix interface is also implemented for \ginkgo, again, in about 350 LOC.

Compared to \hiop's implementations, the \sundials ones are more minimal.
Besides the expected apply and solve method, the matrix and solver interfaces only provide conversions to the \sundials C API, as well as access to the underlying \ginkgo objects.
Similar to \hiop the apply and solve methods work with views on the application data.
However, in contrast to \hiop, no configuration options are available, either as methods or as constructor arguments.
Instead, the \ginkgo interfaces are constructed from existing \ginkgo objects.
This allows users of \sundials to use every possible \ginkgo configuration option, without adding a burden to the interface by translating these options into \sundials options.
Thus, users profit from \ginkgo's flexibility, while the maintenance effort on \sundials is minimal.

\subsection{\opencarp}
\label{sec:cases:opencarp}

\opencarp is a simulator for cardiac electrophysiology.
It provides tools for a wide range of simulations with different physical models.
The simulations are discretized with finite element discretization, which, in the end, leads to solving linear systems.
Through integrating \ginkgo, \opencarp is capable of offloading the linear systems to GPUs~\cite{2022_OpenCARP_Ginkgo_HAPM_Workshop}.

Through the integration of \ginkgo, \opencarp transitioned from a tight coupling to a loose coupling.
Before \ginkgo was used, \opencarp solely relied on \petsc for their linear algebra need.
As a consequence, \opencarp was tightly coupled with \petsc, using their data types throughout their codebase.
Components that are not related to linear algebra, for example the physics module, contained hard dependencies on \petsc\footnote{See for example the file \texttt{physics/electrics.cc} in the release \texttt{v11.0}}.
These dependencies had to be weakened for \ginkgo to be utilized.

The transition to a loose coupling introduced abstract interfaces for vectors, matrices, and solvers.
This required a large-scale effort.
About 8000 LOC were changed in the initial PR\footnote{\url{https://git.opencarp.org/openCARP/openCARP/-/merge_requests/79}} due to the dependency on \petsc in many modules.
Since \opencarp doesn't have a native vector or matrix type like \hiop or \sundials, the interfaces provide a multitude of functionality to cover the previous usage.
Most essential are the typical BLAS level 1 and 2 type operations.
Additionally, methods to help the assembly are present.
Still, even with these additions, the resulting interfaces are relatively small compared to the initial effort, only approximately 300-500 LOC per interface were necessary.

Now, the \opencarp codebase has become more sustainable.
The linear algebra backend implementations are properly separated from other modules, and their dependencies are restricted to the linear algebra module.
It has become possible to easily add other implementations for this component.
As both the \petsc and \ginkgo implementation have a manageable size, 1600 lines, and 1000 lines respectively, the expected effort for additional implementations would again be of moderate size.

\subsection{\openfoam}
\label{sec:cases:openfoam}
\openfoam (Open Field Operation And Manipulation) is a C++ framework for the development of numerical solvers and pre-/post-processing utilities for continuum mechanics problems, especially computational fluid dynamics (CFD)~\cite{jasak2007openfoam}. 
While various efforts were made to provide GPU support to \openfoam no commonly accepted approach exists~\cite{olenik2024Towards} to this date.
One approach is to offload certain parts, e.~g. the linear solver, to GPUs via dedicated plugins.
These plugins are separate from \openfoam.
Their respective libraries are loaded at runtime to provide access to the additional functionality.
Thus, \openfoam interacts through loose coupling with the plugins, while the plugins themselves are implemented using tight coupling.
The \openfoam~\ginkgo~Layer (\ogl)~\cite{olenik2024Towards}\footnote{\url{https://github.com/hpsim/OGL/}} is one of the plugins, providing linear solvers from the \ginkgo library.

\ogl functionality is split into three parts.
First, it provides a mapping between \openfoam and \ginkgo data structures, especially for sparse matrices.
For these Ginkgo data structures, \ogl has to ensure that they persist through the whole simulation.
Lastly, the \ginkgo solvers and preconditioners can be configured at runtime through \ogl.
All combined, \ogl allows \openfoam to efficiently use the platform portable solvers of \ginkgo. 

Since \ogl is a standalone library, it needs a considerably larger effort than the previously discussed integrations.
The library consists roughly of 7000 lines of C++ code.
Currently, the largest part of that is handling the object persistency.
This requires knowledge of the underlying object storage as defined in Ginkgo, which is directly exposed in \ginkgo API.
Without tight coupling, more complex interfaces and hierarchies would be required to describe the storage dynamically for different libraries.
In general, the type hierarchy is shallow, since very little abstraction is necessary.

\section{Summary}
\label{sec:summary}

Many papers discuss the functionality gain that library integrations offer to application software.
Although these gains are the reason why the integration is considered in the first place, the consequences on the development effort for the application are rarely discussed.
This paper offers a discussion on these kinds of impacts.
By the example of the library \ginkgo and its integration into multiple applications, the effects on application software are highlighted.

While \ginkgo offered each application clear gains, e.g. in the form of a platform portable linear algebra component or specialized batched solvers, those came at a cost on the application side.
The \ginkgo integration had to be developed and needed to be maintained, which occupies valuable application developer time.
But under the right circumstances, the integration effort is minimized.
The key factor here is the loose coupling through well-defined interfaces on the application side.
Applications that pursued this approach could integrate \ginkgo in less than 500 LOC per interface, while otherwise a large effort was required by the application, either to switch to loose coupling or build the tight coupling from the ground up.

For the examples given here, loose coupling offered the best approach for application developers.
However, the examples only represent a very limited subset.
Further evaluation of different integration approaches is necessary to provide both library and application developers with concise guidelines.
Another aspect to further evaluate is the extension to other library types.
While this paper only considers linear algebra libraries, the integration of libraries with different focuses might lead to other conclusions.

\begin{acknowledgement}
This work was supported by the “Impuls und Vernetzungsfond of the Helmholtz Association” under grant VH-NG-1241. This work was supported by the European High-Performance Computing Joint Undertaking EuroHPC under grant agreement No 955495 (MICROCARD). Further, it was supported by BMBF through the SCALEXA initiative and the ExaSim (16ME0676K) and PDExa (16ME0641) projects. This research was also supported by the Exascale Computing Project (17-SC-20-SC), a collaborative effort of the U.S. Department of Energy Office of Science and the National Nuclear Security Administration.
\end{acknowledgement}

%
%
%

\bibliographystyle{spmpsci}
\bibliography{ref}

\end{document}